# From powder to part: influence of virgin and recovered Inconel 625 powders on the DED-LP processability, microstructure and mechanical properties

R. DELOFFRE[1,2], L. HERAUD[2], and J. LARTIGAU[1]

[1] Univ. Bordeaux, ESTIA Institute of Technology, F-64210 Bidart, FR
[2] Arts et Métiers Institute of Technology, CNRS, Bordeaux INP, I2M, UMR 5295, F-33400 Talence, FR

*Abstract* - **The reuse of metal powders in directed energy deposition using laser powder offers promising sustainability benefits for additive manufacturing, yet its impact on part quality remains unknown. This study investigates the influence of powder reuse on the directed energy deposition process of Inconel 625. Virgin powder was first characterized in terms of flowability, morphology, size and chemical composition, then used to manufacture parts under fixed process parameters. These parts were analysed through two configurations (as-built and hot isostatically pressed) to assess microstructural evolution and mechanical performance. Results show that hot isostatic pressing reduces porosity and promotes grain recrystallization, enhancing ductility while decreasing tensile strength. Recovered powders from one manufacturing cycle, were also characterized. Particle size distribution and morphology remained stable, interstitial contents, particularly oxygen, increased, potentially affecting laser-material interaction and melt pool behaviour in future production. These findings highlight the importance of evaluating powder reuse through a full-process approach, from powder to final part. Future work will focus on the tensile and fatigue performance of parts produced with recovered powders.**

## 1. Introduction

Directed Energy Deposition – Laser Powder (DED-LP) process is recognized as a key additive manufacturing (AM) technology for the repair, remanufacturing, and production of large and complex metallic parts.

However, DED-LP suffers from a low powder utilization efficiency. Depending on process parameters, up to 60% of the projected powder may be unfused and fall around the part and onto the substrate [1]. This unfused powder is often excluded from the production cycle due to potential alterations in chemical composition, particle morphology, size, or oxidation state. This situation challenges the commonly cited advantages of AM in terms of material savings and raises environmental, economic, and industrial issues.

Therefore, collecting and reusing non-melted powders appears as a promising strategy to enhance the sustainability of the process. While powder bed fusion technologies have been extensively studied, the reuse of powder in DED-LP for nickel-based superalloys such as Inconel 625 (IN625) remains insufficiently documented [8].

In this context, the present work adopts a systematic approach based on a three-part framework: (i) powder characterization, (ii) manufacturing using DED-LP, and (iii) manufactured parts characterization, both in terms of microstructure and mechanical performance.

This methodology is initially applied to virgin IN625 powder (VP), establishing a reference for future comparisons with powder recovered from a first DED-LP production cycle (RP1).

## 2. Virgin powder characterization

The VP was produced by plasma atomization with a nominal particle size distribution (PSD) of 45-90 µm.

Flowability was measured using a Hall flowmeter funnel, the average of three measurements was 12.34 s/50 cm³.

Powder morphology is predominantly spherical, with an average circularity of 0.90, and sphericity of 0.85.

PSD analysis was performed using dynamic image analysis. The distribution was narrow and centered within the expected range ($D_{10}$ = 50.5 µm, $D_{50}$ = 62.3 µm, $D_{90}$ = 80.1 µm).

Chemical composition was measured by Inductively Coupled Plasma Optical Emission Spectroscopy for the main alloying elements, and by combustion/fusion under inert gas for interstitial elements (Tab. I). The measured composition complies with the standard IN625 specification.

TABLE I. INTERSTITIAL ELEMENT CONTENT OF VIRGIN POWDER

| Element | O | N | H | C | S |
|---|---|---|---|---|---|
| Content [ppm] | 61 | 111 | 1,5 | 67 | 0,7 |

## 3. DED-LP manufacturing process

The manufacturing process was carried out using a five axis DED-LP system, equipped with a 2-kW continuous wave fiber laser. A nozzle with a 0.695 mm diameter was selected to produce narrow beads (~1 mm width). The resulting laser spot diameter at focus point was 1.2 mm. No inerting chamber was used due to IN625's low oxidation sensitivity. Instead, local shielding was provided by argon delivered through the carrier and shaping gas at the nozzle. Two net-shape specimen types were manufactured and machined for mechanical testing:

- Tensile specimens: parallelepiped geometry, deposited with a back-and-forth scanning strategy.
- Fatigue specimens: cylindrical geometry, using a similar scanning strategy with a 45° rotation between successive layers.

A fixed parameter set (laser power: 583 W, scanning speed: 2250 mm/min, powder feed rate: 7 g/min) was applied to all builds. This set was defined through a preliminary parametric study to ensure dimensional stability, minimize porosity, and achieve proper track overlap. Despite this optimization, residual porosities - up to 60 µm - were observed in the as-built (AB) condition, motivating the application of hot isostatic pressing (HIP) to assess its impact on microstructure, tensile properties, and fatigue behavior. Future work will evaluate both sieved (RP1-S) and unsieved (RP1-No-S) recovered powders using the same experimental framework.

## 4. Virgin powder parts characterization

The mechanical and microstructural characterizations were investigated on parts manufactured from VP in both AB and HIPed conditions.

### *4.1. Methods*

For metallographic analysis, specimens were sectioned, cold-mounted, mechanically polished, and observed using optical microscopy and scanning electron microscopy (SEM). Room-temperature quasi-static tensile tests were performed using a video extensometer on five tensile specimens in each condition.

### 4.2. *Microstructure*

In the AB condition, the material exhibits a bimodal microstructure composed of elongated columnar grains aligned with the build direction, and fine equiaxed grains located at the bottom of melt pools. EBSD analysis revealed a slight crystallographic texture with a preferred <101> orientation along the deposition axis. This results from the layer-by-layer nature of the DED-LP process, which imposes a strong thermal gradient along the build direction. Subsurface metallurgical defects were observed, including porosities up to 60 µm.

After HIP treatment, the microstructure becomes more homogeneous, featuring large, twinned, and fully equiaxed grains and an average grain size of 52.1 µm. In addition, melt pool boundaries are no longer visible. Porosity is significantly reduced in both pores size and density.

### 4.3. *Mechanical properties*

Tensile properties at room temperature indicate high mechanical performance. In the AB condition, specimens reach a yield strength of 505.2 ± 4.4 MPa, ultimate tensile strength of 841.2 ± 7.2 MPa, and elongation of 50.4 ± 2.1 % (Fig. 1).

Following HIP, the yield strength decreases to 336.6 ± 5.9 MPa, the UTS slightly reduces to 775.8 ± 6.1 MPa, while ductility improves to 56.3 ± 0.8 % (Fig. 1). This evolution is consistent with stress relief and mainly attributed to microstructural homogenization, including grain growth and the decrease in dislocation density.

These results suggest that HIP treatment does not lead to significant enhancement in tensile properties. However, the reduction in internal metallurgical defects could have a significant impact on fatigue behaviour, which remains to be assessed.

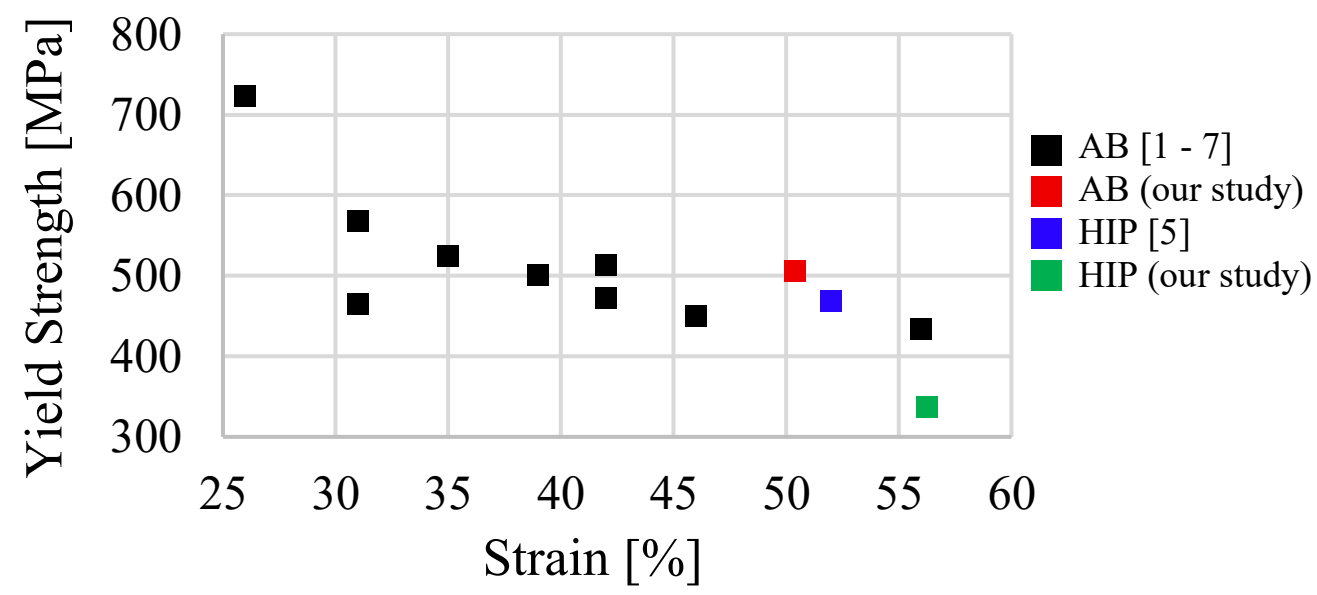

Figure 1. IN625 DED-LP tensile behaviour: AB vs HIP states [1 – 7].

## 5. Recovered powders characterization

To assess powder reuse, RP1 was analysed in two conditions – as recovered and unsieved (RP1-No-S), and after sieving (RP1-S) – and compared to VP.

At first order, no significant degradation was observed compared to VP:

- PSD remain stable with similar percentile diameters, indicating limited agglomeration or satellites formation after one manufacturing cycle.
- Morphological features remain unchanged, with mostly spherical particles showing localized morphological defects when observed under SEM.
- Flowability is preserved (12.19 s/50cm$^3$ for RP1-No-S), and even slightly improved after sieving (11.61 s/50cm$^3$).
- Chemical composition shows very slight variations.

The main difference lies in the increased contents of oxygen, carbon, and hydrogen (Tab. II). In particular, the rise in oxygen content in RP1 powders may result from particle-atmosphere interactions of particles that remain hot after laser exposure, thereby promoting surface oxidation.

TABLE II. INTERSTITIAL ELEMENT CONTENT OF RECOVERED POWDERS

| Element content [ppm] | O | N | H | C | S |
|---|---|---|---|---|---|
| RP1-No-S | 368 | 113 | 3.1 | 80 | 0.5 |
| RP1-S | 365 | 113 | 3.1 | 79 | 0.5 |

A direct comparison between RP1-No-S and RP1-S indicates that sieving has no significant effect on powder characteristics except for flowability. Interstitial element contents remain unchanged, suggesting that sieving alone is not sufficient to restore chemical integrity of recovered powders.

These preliminary analyses raise important questions about the potential influence of powder reuse on DED-LP process, resulting microstructure, and mechanical performance. For instance, higher oxygen content may enhance laser-energy absorption, potentially altering melt pool behaviour [8]. These chemical evolutions highlight the need for further investigation across the full chain – from process to microstructure and mechanical properties.

## 6. Conclusion

This study provides a baseline characterization of virgin and recovered IN625 powders for DED-LP additive manufacturing process. The results show that, both morphologically and in terms of PSD, RP1 powders remain highly comparable to VP, regardless of sieving. However, the manufacturing cycle induces a notable increase in interstitial elements (O, C, H), which may influence the laser-material interaction and melt pool behaviour for future parts.

The elevated oxygen content raises concerns regarding possible changes in processability of recovered powder: energy absorption, melt pool dynamics, and the formation of metallurgical defects. These findings highlight the importance of continuing the investigation encompassing the entire additive manufacturing chain – from powder to part.